\documentclass[11pt]{article}
\usepackage{blois,epsfig}

\def\be{\begin{equation}}
\def\ee{\end{equation}}
\def\bea{\begin{eqnarray}}
\def\eea{\end{eqnarray}}

\def\gtwid{\mathrel{\raise.3ex\hbox{$>$\kern-1.05em\lower1ex\hbox{
$\sim$}}}}
\def\ltwid{\mathrel{\raise.3ex\hbox{$<$\kern-1.05em\lower1ex\hbox{
$\sim$}}}}

\begin{document}
\vspace*{4cm}
\title{AN ARGUMENT THAT THE DARK MATTER IS AXIONS}

\author{ P. SIKIVIE }

\address{Department of Physics, University of Florida,\\
Gainesville, FL 32611, USA}

\maketitle

\abstracts{ An argument is presented that the dark matter is axions, 
at least in part.  It has three steps.  First, axions behave differently 
from the other forms of cold dark matter because they form a rethermalizing
Bose-Einstein condensate (BEC).  Second, there is a tool to distinguish 
axion BEC from the other dark matter candidates on the basis of observation, 
namely the study of the inner caustics of galactic halos.  Third, the 
observational evidence for caustic rings of dark matter is consistent 
in every aspect with axion BEC, but not with the other proposed forms
of dark matter.}

One of the outstanding problems in science today is the identity
of the dark matter of the universe \cite{PDM}.  The existence of
dark matter is implied by a large number of observations, including
the dynamics of galaxy clusters, the rotation curves of individual
galaxies, the abundances of light elements, gravitational lensing,
and the anisotropies of the cosmic microwave background radiation.
The energy density fraction of the universe in dark matter is 23\%.
The dark matter must be non-baryonic, cold and collisionless.  Particles
with the required properties are referred to as `cold dark matter' (CDM).
The leading CDM candidates are weakly interacting massive particles
(WIMPs) with mass in the 100 GeV range, axions with mass in the
$10^{-5}$ eV range, and sterile neutrinos with mass in the keV
range.

Based on work by my collaborators and I over the past three or 
four years, I have come to the conclusion that the dark matter is 
axions, at least in part. The starting point is that cold dark matter 
axions thermalize as a result of their gravitational self-interactions
\cite{CABEC}.  When they thermalize, they form a Bose-Einstein condensate.  
It may seem surprising that axions thermalize as a result of their 
gravitational self-interactions since gravitational interactions 
among particles are usually thought to be negligible.  Gravitational 
interactions among particles are in fact almost always negligible but 
cold dark matter axions are an exception because the axions occupy in 
huge numbers a small number of states (the typical quantum state 
occupation number is $10^{61}$) and those states have enormous 
correlation lengths (of order parsec to Gpc, today). 

Let us call $\Gamma = 1/\tau$ the axion thermalization rate.  On time 
scales short compared to $\tau$, cold dark matter axions form a degenerate 
Bose gas described by a classical field equation.  Their behavior is then
indistinguishable from that of ordinary CDM except on length scales that 
are too short ($10^{14}$ cm or so) to be of observational interest. On 
times scales large compared to $\tau$, cold dark matter axions thermalize.  
The thermalization of a degenerate Bose gas is a quantum-mechanical entropy 
generating process, not described by classical field equations.  On time 
scales larger than $\tau$ the axion state, i.e. the state that most axions 
are in, tracks the lowest energy state available to them.  The behaviour of 
such a rethermalizing axion BEC is different from that of ordinary CDM and 
the differences are observable.

The thermalization of cold dark matter axions is discussed in detail 
in ref. \cite{therm}.  It is found there that rethermalization of the 
axion BEC by gravitational self-interactions is sufficiently fast that 
the axions that are about to fall into a galactic potential well almost 
all go to the lowest energy state consistent with the angular momentum 
they have acquired from tidal torquing.  That state is one of net overall 
rotation, implying 
$\vec{\nabla} \times \vec{v} \neq 0$ where $\vec{v}(\vec{x}, t)$ is 
the velocity field of the infalling dark matter.  In contrast, 
ordinary cold dark matter (e.g. WIMPs and sterile neutrinos) 
falls in with an irrotational velocity field, 
$\vec{\nabla} \times \vec{v} = 0$.  The inner caustics of galactic 
halos are different in the two cases.  If the dark matter falls 
in with net overall rotation, the inner caustics are rings whose 
cross-section is a section of the elliptic umbilic ($D_{-4}$)
catastrophe, called caustic rings for short \cite{crdm}.  If
the velocity field of the infalling particles is irrotational, the 
inner caustics have a 'tent-like' structure which is described in 
detail in ref. \cite{inner} and which is quite distinct from that 
of caustic rings.  Evidence was found for caustic rings of dark 
matter.  The evidence is summarized in ref. \cite{MWhalo}.  It is 
shown in ref. \cite{case} that the evidence for caustic rings of 
dark matter is precisely and in all respects consistent with the 
predictions of a rethermalizing axion BEC.

The above is the gist of the argument.  It is elaborated further in 
the three sections below.  A few comments may be in order.  One question 
is: what fraction of the dark matter must be axions to justify the evidence 
for caustic rings. We  hope to comment on this soon.  Another question is: 
to what extent does the evidence for caustic rings require the dark matter 
to be QCD axions \cite{axion}, as opposed to some other kind of axion-like 
particle(s).  The evidence requires that a sizable fraction of the dark 
matter be identical bosons, whose number is conserved on cosmological 
time scales, and which are sufficiently cold and thermalize sufficiently 
fast that they form a BEC.  Furthermore the BEC must rethermalize 
sufficiently fast that the particles go to a state of net overall 
rotation as they are about to fall into galactic potential wells.  
It happens that the QCD axion with mass of order $10^{-5}$ eV has all 
these properties and since it solves in addition the strong CP problem 
of the Standard Model of elementary particles, it is reasonable to assume 
that the dark matter is in fact QCD axions.  However, there are many 
axion-like particles \cite{cada} that can equally well reproduce the 
evidence for caustics rings.  Furthermore, whether or not the particle in 
question is the QCD axion, the prediction of Bose-Einstein condensation 
and subsequent caustic ring formation is rather insensitive to the 
particle mass and therefore does not provide a good guide to it.  
The axion is being searched for as a constituent of the Milky Way 
halo \cite{ADMX}, as a particle radiated by the Sun \cite{helio}, 
and in experiments that convert photons to axions and axions back 
to photons behind a wall \cite{SLW}.  

Finally, many authors have proposed \cite{dmBEC,Rind} that the dark 
matter is a Bose-Einstein condensate of particles with mass of order 
$10^{-21}$ eV or less.  When the mass is that small, the dark matter 
BEC behaves differently from CDM on scales of observational interest 
as a result of the tendency of the BEC to delocalize.  Due to the 
Heisenberg uncertainty principle, a BEC has Jeans' length 
\cite{Ruff,CABEC,Chav} 
\begin{equation}
\ell_{\rm J} = (16 \pi G \rho m^2)^{-{1 \over 4}}
= 1.02 \cdot 10^{14}~{\rm cm}
\left({10^{-5}~{\rm eV} \over m}\right)^{1 \over 2}
\left({10^{-29}~{\rm g/cm^3} \over \rho}\right)^{1 \over 4}~\ ,
\label{Jeans}
\end{equation}
where $\rho$ is the BEC density and $m$ the constituent particle 
mass.  As mentioned earlier, this length scale is unobservably 
small in the QCD axion case.  In contrast, when $m \sim 10^{-21}$ 
eV, the Jeans' length is of order 3 kpc and has implications for 
observation.  It leads to a suppression of the dark matter 
density near the galactic center.  This may be a remedy for 
the excessive concentration of dark matter near galactic 
centers seen in numerical simulatons of structure formation 
with ordinary CDM \cite{conc}.

\section{Bose-Einstein condensation of cold dark matter axions}

Shortly after the Standard Model of elementary particles was
established, the axion was postulated \cite{axion} to explain
why the strong interactions conserve the discrete symmetries P
and CP.  For our purposes the action density for the axion field
$\varphi(x)$ may be taken to be
\begin{equation}
{\cal L}_a = - {1 \over 2} \partial_\mu \varphi \partial^\mu \varphi
- {1 \over 2} m^2 \varphi^2 + {\lambda \over 4!} \varphi^4 - ...
\label{lag}
\end{equation}
where $m$ is the axion mass.  The self-coupling strength is \cite{CABEC}
\begin{equation}
\lambda = {m^2 \over f^2}~{m_d^3 + m_u^3 \over (m_d + m_u)^3}
\simeq 0.35~{m^2 \over f^2}
\label{self}
\end{equation}
in terms of the axion decay constant $f$ and the masses $m_u$ and
$m_d$ of the up and down quarks.  In Eq.~(\ref{lag}), the dots  
represent higher order axion self-interactions and interactions 
of the axion with other particles.  All axion couplings and the
axion mass
\begin{equation}
m \simeq 6 \cdot 10^{-6}~{\rm eV}~{10^{12}~{\rm GeV} \over f}
\label{mass}
\end{equation}
are inversely proportional to $f$. $f$ was first thought to be of order   
the electroweak scale, but its value is in fact arbitrary \cite{invis}. 
However, the combined limits from unsuccessful searches in particle
and nuclear physics experiments and from stellar evolution require
$f \gtwid 3 \cdot 10^9$ GeV \cite{axrev}.

Furthermore, an upper limit $f \ltwid 10^{12}$ GeV is provided by
cosmology because light axions are abundantly produced during the QCD
phase transition \cite{axdm}.  In spite of their very small mass, these
axions are a form of cold dark matter.  Indeed, their average momentum
at the QCD epoch is not of order the temperature (GeV) but of order the
Hubble expansion rate ($3 \cdot 10^{-9}$ eV) then.  In case inflation
occurs after the Peccei-Quinn phase transition their average momentum is
even smaller because the axion field gets homogenized during inflation.
In addition to the cold axion population, there is a thermal axion 
population with average momentum of order the temperature.

The non-perturbative QCD effects that give the axion its mass turn on
at a temperature of order 1 GeV.  The critical time, defined by
$m(t_1) t_1 = 1$, is $t_1 \simeq 2 \cdot 10^{-7}~{\rm sec}~
(f / 10^{12}~{\rm GeV})^{1 \over 3}$.  Cold axions are the quanta
of oscillation of the axion field that result from the turn on of
the axion mass.  They have number density
\begin{equation}
n(t) \sim {4 \cdot 10^{47} \over {\rm cm}^3}~
\left({f \over 10^{12}~{\rm GeV}}\right)^{5 \over 3}
\left({a(t_1) \over a(t)}\right)^3
\label{numden}
\end{equation}
where $a(t)$ is the cosmological scale factor.  Because the axion 
momenta are of order ${1 \over t_1}$ at time $t_1$ and vary with   
time as $a(t)^{-1}$, the velocity dispersion of cold axions is   
\begin{equation}
\delta v (t) \sim {1 \over m t_1}~{a(t_1) \over a(t)}
\label{veldis}
\end{equation}
{\it if} each axion remains in whatever state it is in, i.e. if axion
interactions are negligible.  The average state occupation number 
of cold axions is then
\begin{equation}
{\cal N} \sim~ n~{(2 \pi)^3 \over {4 \pi \over 3} (m \delta v)^3}   
\sim 10^{61}~\left({f \over 10^{12}~{\rm GeV}}\right)^{8 \over 3}~~\ .
\label{occnum}
\end{equation}
That ${\cal N}$ is much larger than one tells us that the effective
temperature of cold axions is much smaller than the critical temperature
for Bose-Einstein condensation.

Bose-Einstein condensation may be briefly described as follows: if 
identical bosonic particles are highly condensed in phase space, if 
their total number is conserved and if they thermalize, most of them go 
to the lowest energy available state.  The condensing particles do so 
because, by yielding their energy to the remaining non-condensed particles, 
the total entropy is increased.  Eq.~(\ref{occnum}) tells us that the first
condition is overwhelmingly satisfied.  The second condition is also
satisfied because all axion number violating processes, such as their
decay to two photons, occur on time scales vastly longer than the
age of the universe.  The only condition for axion BEC that is not
manifestly satisfied is thermal equilibrium.

Axions are in thermal equilibrium if their relaxation rate $\Gamma$ is
large compared to the Hubble expansion rate $H(t) = {1 \over 2t}$.  The
relaxation rate $\Gamma$ is given in the {\it particle kinetic} regime by
\begin{equation}
\Gamma \sim n~\sigma~\delta v~{\cal N}
\label{parkin}
\end{equation}
where $\sigma$ is the relevant scattering cross-section.  The particle
kinetic regime is defined by the condition $\Gamma << \delta E$ where
$\delta E$ is the energy dispersion of the particles.  Cold dark matter
axions are however mostly in the opposite regime: $\Gamma >> \delta E$ 
which we call the {\it condensed regime}.  Thermalization in the condensed
regime is discussed in detail in our recent paper \cite{therm}.  We find
that the relaxation rate of cold axions through their $\lambda \phi^4$
self-interactions is of order \cite{CABEC,therm}
\begin{equation}
\Gamma_\lambda  \sim {1 \over 4} \lambda~n~m^{-2}~~\ .
\label{rate2} 
\end{equation}
$\Gamma_\lambda(t)/H(t)$ is of order one at time $t_1$ but
decreases as $t~a(t)^{-3} \propto a(t)^{-1}$ afterwards, 
implying that cold axions briefly thermalize as a result of 
their $\lambda \phi^4$ interactions when they are first produced 
during the QCD phase transition but, after this brief period of 
thermalization, the axions are decoupled again.

However the axions rethermalize later as a result of their
gravitational self-interactions.  Their relaxation rate by
gravitational interactions is of order \cite{CABEC,therm}
\begin{equation}
\Gamma_{\rm g} \sim 4 \pi G~n~m^2~l^2
\label{rate3}
\end{equation}
where $l \sim (m \delta v)^{-1}$ is their correlation length.
$\Gamma_{\rm g}(t)/H(t)$ is of order
$5 \cdot 10^{-7}(f/10^{12}~{\rm GeV})^{2 \over 3}$
at time $t_1$ but grows as $t a^{-1}(t) \propto a(t)$.
Thus gravitational interactions cause the axions to thermalize
and form a BEC when the photon temperature is of order
500 eV~$(f/10^{12}~{\rm GeV})^{1 \over 2}$.  Bose-Einstein
condensation causes the axion correlation length to grow 
until it becomes of order the horizon.  The growth in the 
correlation length causes the thermalization to accelerate;  
see Eq.~(\ref{rate3}).  When $l$ is some fraction of $t$, 
$\Gamma_g(t)/H(t) \propto a(t)^{-3}t^3$, implying that 
thermalization occurs on ever shorter time scales compared 
to the Hubble time.  The question is now whether axion BEC 
has implications for observation. 

\section{Dark matter caustics}

The study of the inner caustics of galactic halos \cite{crdm,inner}  
provides a useful tool.  An isolated galaxy like our own accretes
the dark matter particles surrounding it.  Cold collisionless particles
falling in and out of a gravitational potential well necessarily form an
inner caustic, i.e. a surface of high density, which may be thought of as
the envelope of the particle trajectories near their closest approach to
the center.  The density diverges at caustics in the limit where the
velocity dispersion of the dark matter particles vanishes.  Because
the accreted dark matter falls in and out of the galactic gravitational
potential well many times, there is a set of inner caustics.  In addition,
there is a set of outer caustics, one for each outflow as it reaches its
maximum radius before falling back in.  We exploit the catastrophe
structure and spatial distribution of the inner caustics of isolated
disk galaxies.

The catastrophe structure of the inner caustics depends mainly on the angular
momentum distribution of the infalling particles \cite{inner}.  There are
two contrasting cases to consider.  In the first case, the angular momentum
distribution is characterized by `net overall rotation';  in the second case,
by irrotational flow.  The archetypical example of net overall rotation is
instantaneous rigid rotation on the turnaround sphere.  The turnaround sphere
is defined as the locus of particles which have zero radial velocity with
respect to the galactic center for the first time, their outward Hubble flow
having just been arrested by the gravitational pull of the galaxy.  The
present turnaround radius of the Milky Way is of order 2 Mpc.  Net overall
rotation implies that the velocity field has a curl, $\vec{\nabla} \times
\vec{v} \neq 0$.  The corresponding inner caustic is a closed tube whose
cross-section is a section of the elliptic umbilic ($D_{-4})$ catastrophe
\cite{crdm,inner}.  We call it a `caustic ring', or `tricusp ring' in
reference to its shape.  In the case of irrotational flow,
$\vec{\nabla} \times \vec{v} = 0$, the inner caustic has a tent-like
structure quite distinct from a caustic ring.  Both types of inner
caustic are described in detail in ref.\cite{inner}.

If a galactic halo has net overall rotation and its time evolution
is self-similar, the radii of its caustic rings are predicted in
terms of a single parameter, called $j_{\rm max}$.  Self-similarity
means that the entire phase space structure of the halo is time
independent except for a rescaling of all distances by $R(t)$, all
velocities by $R(t)/t$ and all densities by $1/t^2$. \cite{FG,B,STW,MWhalo}
For definiteness, $R(t)$ is taken to be the turnaround radius at
time $t$.  If the initial overdensity around which the halo forms has
a power law profile \cite{FG}
\begin{equation}
{\delta M_i \over M_i} \propto \left({1 \over M_i}\right)^\epsilon~~~\ ,  
\label{inov}
\end{equation}
where $M_i$ and $\delta M_i$ are respectively the mass and excess mass within 
an initial radius $r_i$, then its subsequent evolution is self-similar with
$R(t) \propto t^{{2 \over 3} + {2 \over 9 \epsilon}}$.     
In an average sense, $\epsilon$ is related to the slope of the evolved
power spectrum of density perturbations on galaxy scales.  The observed
power spectrum implies that $\epsilon$ is in the range 0.25 to 0.35
\cite{STW}.  The prediction for the caustic ring radii is ($n$ = 1, 2, 3,
.. ) \cite{crdm,MWhalo}
\begin{equation}
a_n \simeq {{\rm 40~kpc} \over n}~
\left({v_{\rm rot} \over 220~{\rm km/s}}\right)~
\left({j_{\rm max} \over 0.18}\right)
\label{crr}
\end{equation}
where $v_{\rm rot}$ is the galactic rotation velocity.
Eq.(~\ref{crr}) is for $\epsilon = 0.3$.  The $a_n$ have a
small $\epsilon$ dependence.  However, the $a_n \propto 1/n$
approximate behavior holds for all $\epsilon$ in the range 0.25
and 0.35.

Observational evidence for caustic rings of dark matter with the radii
predicted by Eq.~(\ref{crr}) was found in: the statistical distribution   
of bumps in a set of 32 extended and well-measured galactic rotation
curves, the distribution of bumps in the rotation curve of the Milky Way,
the appearance of a triangular feature in the IRAS map of the Milky Way
in the precise direction tangent to the nearest caustic ring, and the  
existence of a ring of stars at the location of the second ($n$ = 2)
caustic ring in the Milky Way.  The observational evidence for caustic   
rings of dark matter is summarized in ref. \cite{MWhalo}.  The recent
improved measurement \cite{Chemin} of the rotation curve of our nearest  
large neighbor, the Andromeda galaxy, provides new evidence.  The new
rotation curve shows three prominent bumps at radii 10 kpc, 15 kpc and
29 kpc, whose ratios accord with Eq.~(\ref{crr}).

To reproduce the evidence for caustic rings of dark matter, the specific
angular momentum distribution on the turnaround sphere should be given by
\begin{equation}
\vec{\ell}(\hat{n},t) =
j_{\rm max}~\hat{n} \times (\hat{z} \times \hat{n})
~{R(t)^2 \over t}
\label{samd}
\end{equation}
where $\hat{n}$ is the unit vector pointing to a position on the turnaround
sphere,  $\hat{z}$ is the axis of rotation and $j_{\rm max}$ is the parameter
that appears in Eq.~(\ref{crr}).  Eq.~(\ref{samd}) states that the turnaround
sphere at time $t$ rotates with angular velocity
$\vec{\omega} = {j_{\rm max} \over t} \hat{z}$.  Each property  of the angular
momentum distribution (\ref{samd}) maps onto an observable property of the inner
caustics:  net overall rotation causes the inner caustics to be rings, the value
of $j_{\rm max}$ determines their overall size, and the time dependence given
in Eq.~(\ref{samd}) is responsible for $a_n \propto 1/n$.  We now show that each
of these three properties follows from the assumption that the infalling dark
matter is a rethermalizing axion BEC.

\section{Three successes}

\subsection{Magnitude of angular momentum}

We make the standard assumption that the angular momentum of a galaxy 
is due to the tidal torque applied to it by nearby protogalaxies early 
on when density perturbations are still small and protogalaxies close 
to one another \cite{ttt}.  The amount of angular momentum galaxies
typically acquire by tidal torquing can be reliably estimated by numerical
simulation because it does not depend on any small feature of the mass 
configuration,  so that the resolution of present simulations is not
an issue in this case.  The dimensionless angular momentum parameter
\begin{equation}
\lambda \equiv {L |E|^{1 \over 2} \over G M^{5 \over 2}}~~\ ,  
\label{lambda}
\end{equation}
where $G$ is Newton's gravitational constant, $L$ is the angular
momentum of the galaxy, $M$ its mass and $E$ its net mechanical
(kinetic plus gravitational potential) energy, was found \cite{Efst}
to have median value 0.05.  In the caustic ring model the magnitude of 
angular momentum is given by $j_{\rm max}$.  The evidence for caustic 
rings implies that the $j_{\rm max}$-distribution is peaked at 
$j_{\rm max} \simeq$ 0.18.  The relationship between $j_{\rm max}$ 
and $\lambda$ is \cite{case}      
\begin{equation}
\lambda = \sqrt{6 \over 5 - 3 \epsilon}~
{8 \over 10 + 3 \epsilon}~
{1 \over \pi}~j_{\rm max}~~~\ .
\label{rel}
\end{equation}
For $j_{\rm max}$ =0.18 and $\epsilon$ in the range 0.25 to 0.35,
Eq.~(\ref{rel}) implies $\lambda$ =0.051. The excellent agreement
between $j_{\rm max}$ and $\lambda$ gives further credence to the
caustic ring model.  Indeed if the evidence for caustic rings were
incorrectly interpreted, there would be no reason for it to produce
a value of $j_{\rm max}$ consistent with $\lambda$.

\subsection{Net overall rotation}

Next we ask whether net overall rotation is an expected
outcome of tidal torquing.  The answer is clearly no if
the dark matter is ordinary CDM. Indeed, the velocity
field of ordinary CDM satisfies
\begin{equation}
{d \vec{v} \over dt}(\vec{r}, t) =
{\partial \vec{v} \over \partial t}(\vec{r}, t) +
(\vec{v}(\vec{r}, t) \cdot \vec{\nabla}) \vec{v} (\vec{r}, t)
= - \vec{\nabla} \phi(\vec{r}, t)
\label{cdm}
\end{equation}
where $\phi(\vec{r}, t)$ is the gravitational potential.  The
initial velocity field is irrotational because the expansion
of the universe caused all rotational modes to decay away.
Furthermore, it is easy to show \cite{inner} that if
$\vec{\nabla} \times \vec{v} = 0$ initially, then
Eq.~(\ref{cdm}) implies $\vec{\nabla} \times \vec{v} = 0$
at all later times.  Since net overall rotation requires
$\vec{\nabla} \times \vec{v} \neq 0$, it is inconsistent with
ordinary CDM, such as WIMPs or sterile neutrinos.  If WIMPs or 
sterile neutrinos are the dark matter, the evidence for caustic 
rings, including the agreement between $j_{\rm max}$ and $\lambda$ 
obtained above, is fortuitous.

Axions do not obey Eq.~(\ref{cdm}) because they form a rethermalizing 
BEC \cite{therm}.  By {\it rethermalizing} we mean that the thermalization 
rate is larger than the Hubble rate so that the axion state tracks the 
lowest energy available state.  The compressional (scalar) modes of 
the axion field are unstable and grow as for ordinary CDM, except on 
length scales too small to be of observational interest \cite{CABEC}.  
Unlike ordinary CDM, however, the rotational (vector) modes of the 
axion field exchange angular momentum by gravitational interaction.  
Most axions condense into the state of lowest energy consistent with 
the total angular momentum, say $\vec{L} = L \hat{z}$, acquired by 
tidal torquing at a given time.  To find this state we may use the 
WKB approximation because the angular momentum quantum numbers are very
large, of order $10^{20}$ for a typical galaxy.  The WKB approximation
maps the axion wavefunction onto a flow of classical particles with
the same energy and momentum densities.  It is easy to show that for
given total angular momentum the lowest energy is achieved when the
angular motion is rigid rotation.  Rigid rotation is therefore   
a prediction of tidal torque theory if the dark matter is axions. 

\subsection{Self-similarity}

Because the axion BEC rethermalization rate is larger then the Hubble
rate, most axions go to the lowest energy state consistent with the 
total angular momentum acquired from nearby inhomogeneities by tidal 
torquing.  The time dependence of the specific angular momentum
distribution on the turnaround sphere is then predicted.   Is it
consistent with Eq.~(\ref{samd})?  In particular, is the axis of
rotation constant in time?

Consider a comoving sphere of radius $S(t) = S a(t)$ centered on the
protogalaxy.  As before, $a(t)$ is the cosmological scale factor.  $S$
is taken to be of order but smaller than half the distance to the nearest
protogalaxy of comparable size, say one third of that distance.  The total
torque applied to the volume $V$ of
the sphere is
\begin{equation} \vec{\tau}(t) = \int_{V(t)} d^3r
~\delta\rho(\vec{r}, t)~\vec{r}\times (-\vec{\nabla} \phi(\vec{r}, t))
\label{torq}
\end{equation}
where $\delta\rho(\vec{r}, t) = \rho(\vec{r}, t) - \rho_0(t)$ is
the density perturbation.  $\rho_0(t)$ is the unperturbed density.
In the linear regime of evolution of density perturbations, the
gravitational potential does not depend on time when expressed in
terms of comoving coordinates, i.e.
$\phi(\vec{r} = a(t) \vec{x}, t) = \phi(\vec{x})$.  Moreover
$\delta(\vec{r}, t) \equiv {\delta \rho(\vec{r}, t) \over \rho_0(t)}$
has the form $\delta(\vec{r} = a(t) \vec{x}, t) = a(t) \delta(\vec{x})$.
Hence
\begin{equation} \vec{\tau}(t) = \rho_0(t) a(t)^4 \int_V d^3x
~\delta(\vec{x})~\vec{x} \times (- \vec{\nabla}_x \phi(\vec{x}))~~~\ .
\label{tt}
\end{equation}
Eq.~(\ref{tt}) shows that the direction of the torque is time independent.
Hence the rotation axis is time independent, as in the caustic ring model.
Furthermore, since $\rho_0(t) \propto a(t)^{-3}$,
$\tau(t) \propto a(t) \propto t^{2 \over 3}$
and hence $\ell(t) \propto L(t) \propto t^{5 \over 3}$.  Since      
$R(t) \propto t^{{2 \over 3} + {2 \over 9 \epsilon}}$, tidal
torque theory predicts the time dependence of Eq.~(\ref{samd})
provided $\epsilon = 0.33$. This value of $\epsilon$ is in
the range, $0.25 < \epsilon < 0.35$, predicted by the evolved   
spectrum of density perturbations and supported by the evidence   
for caustic rings.  So the time dependence of the angular momentum
distribution on the turnaround sphere is also consistent with the  
caustic ring model.

In conclusion, the phase space structure of galactic halos predicted
by tidal torque theory, if the dark matter is axions, is precisely
and in all respects that of the caustic ring model proposed earlier
on the basis of observations.  The other dark matter candidates predict
a different, more chaotic phase space structure for galactic halos.
Although the QCD axion is best motivated, a broader class of axion-like
particles behaves in the manner described here.

\section*{Acknowledgments}
I would like to thank the Aspen Center for Physics for its
support (NSF Grant \#1066293) and its hospitality while working 
on this paper.  This work was supported in part by the U.S. 
Department of Energy under grant DE-FG02-97ER41209.


\section*{References}
\begin{thebibliography}{99}

\bibitem{PDM}
For a recent review, see {\it Particle Dark Matter}, edited
by Gianfranco Bertone, Cambridge University Press 2010.

\bibitem{CABEC} 
P. Sikivie and Q. Yang, Phys. Rev. Lett. 103 (2009) 111301.

\bibitem{therm}
O. Erken, P. Sikivie, H. Tam and Q. Yang, Phys. Rev. D85
(2012) 063520.

\bibitem{crdm}
P. Sikivie, Phys. Lett. B432 (1998) 139;
Phys. Rev. D60 (1999) 063501.

\bibitem{inner}
A. Natarajan and P. Sikivie, Phys. Rev. D73 (2006) 023510.

\bibitem{MWhalo} 
L.D. Duffy and P. Sikivie, Phys. Rev. D78 (2008) 063508.

\bibitem{case}
P. Sikivie, Phys. Lett. B695 (2011) 22.

\bibitem{axion}
R. D. Peccei and H. Quinn, Phys. Rev. Lett. 38 (1977) 1440 and Phys.
Rev. D16 (1977) 1791; S. Weinberg, Phys. Rev. Lett. 40 (1978) 223;   
F. Wilczek, Phys. Rev. Lett. 40 (1978) 279.

\bibitem{cada}
For a recent discussion of a broad class of axion-like particles, see:  
P. Arias et al., JCAP 06 (2012) 013.

\bibitem{ADMX}
S.J. Asztalos et al., Phys. Rev. Lett. 104 (2010) 041301, and 
references therein.

\bibitem{helio}
S. Aune et al., Phys. Rev. Lett. 107 (2011) 261302; 
R. Ohta et al., Nucl. Instr. Meth. A670 (2012) 73.

\bibitem{SLW}
K. Ehret et al., Phys. Lett. B689 (2010) 149;
G. Mueller et al., Phys. Rev. D80 (2009) 072004, and 
references therein.

\bibitem{dmBEC}
S.-J. Sin, Phys. Rev. D50 (1994) 3650;
J. Goodman, New Astronomy Reviews 5 (2000) 103;
W. Hu, R. Barkana and A. Gruzinov, Phys. Rev. Lett. 85 (2000) 1158;
E.W. Mielke and J.A. V\'elez P\'erez, Phys. Lett. B671 (2009) 174;
J.-W. Lee and S. Lim, JCAP 1001 (2010) 007;
A. Lundgren, M. Bondarescu, R. Bondarescu and J. Balakrishna,
Ap. J. 715 (2010) L35;
D.J. Marsh and P.G. Ferreira, Phys. Rev. D82 (2010) 103528;
V. Lora et al., JCAP 02 (2012) 011.

\bibitem{Rind}
T. Rindler-Daller and P. Shapiro, MNRAS 422 (2012) 135, and 
references therein.

\bibitem{Ruff}
M. Bianchi, D. Grasso and R. Ruffini, Astron. Astroph. 231 (1990) 301.

\bibitem{Chav}
P.-H. Chavanis, Astroph. J. 537 (2012) A127.

\bibitem{conc}
J.F. Navarro and W. Benz, Ap. J. 380 (1991) 320;
S.D.M. White and J.F. Navarro, MNRAS 265 (1993) 271;
J.F. Navarro and M. Steinmetz, Ap. J. 513 (1999) 555.

\bibitem{invis}
J. Kim, Phys. Rev. Lett. 43 (1979) 103; M. A. Shifman,
A. I. Vainshtein and V. I. Zakharov, Nucl. Phys. B166 (1980) 493;
A. P. Zhitnitskii, Sov. J. Nucl. 31 (1980) 260;  M. Dine,
W. Fischler and M. Srednicki, Phys. Lett. B104 (1981) 199.

\bibitem{axrev}
J.E. Kim, Phys. Rep. 150 (1987) 1;
M.S. Turner, Phys. Rep. 197 (1990) 67;
G.G. Raffelt, Phys. Rep. 198 (1990) 1.

\bibitem{axdm}
J. Preskill, M. Wise and F. Wilczek, Phys. Lett. B120 (1983) 127;
L. Abbott and P. Sikivie, Phys. Lett. B120 (1983) 133;
M. Dine and W. Fischler, Phys. Lett. B120 (1983) 137.

\bibitem{FG} 
J.A. Fillmore and P. Goldreich, Ap. J. 281 (1984) 1.

\bibitem{B}
E. Bertschinger, Ap. J. Suppl. 58 (1985) 39.

\bibitem{STW}
P. Sikivie, I. Tkachev and Y. Wang, Phys. Rev. Lett. 75 (1995) 2911;     
Phys. Rev. D56 (1997) 1863.

\bibitem{Chemin}
L. Chemin, C. Carignan and T. Foster, Ap. J. 705 (2009) 1395.  

\bibitem{ttt}
G. Stromberg, Ap. J. 79 (1934) 460;
F. Hoyle, in {\it Problems of Cosmical Aerodynamics}, ed. by
J.M. Burgers and H.C. van de Hulst, 1949, p195.  Dayton, Ohio:
Central Air Documents Office; P.J.E. Peebles, Ap. J. 155 (1969) 2, 
and Astron. Ap. 11 (1971) 377.

\bibitem{Efst}
G. Efstathiou and B.J.T. Jones, MNRAS 186 (1979) 133;  
J. Barnes and G. Efstathiou, Ap. J. 319 (1987) 575;
B. Cervantes-Sodi et al., Rev. Mex. AA. 34 (2008) 87.

\end{thebibliography}
\end{document}